\documentclass[aps,amsmath,twocolumn,prl,showpacs]{revtex4}

\usepackage{graphicx}
\usepackage{color}

\begin{document}


\title{Spectrum and thermal fluctuations of a microcavity polariton Bose-Einstein condensate}

\author{D. Sarchi}
\email[]{davide.sarchi@epfl.ch}
\affiliation{Institute of Theoretical Physics, Ecole Polytechnique F\'ed\'erale de Lausanne EPFL, CH-1015 Lausanne, Switzerland}
\author{V. Savona}
\affiliation{Institute of Theoretical Physics, Ecole Polytechnique F\'ed\'erale de Lausanne EPFL, CH-1015 Lausanne, Switzerland}

\date{\today}

\begin{abstract}
The Hartree-Fock-Popov theory of interacting Bose particles is developed, for modeling exciton-polaritons in semiconductor
microcavities undergoing Bose-Einstein condensation. A self-consistent
treatment of the linear exciton-photon coupling and of the exciton non-linearity
provides a thermal equilibrium description of the collective excitation spectrum, of the polariton
energy shifts and of the phase diagram. Quantitative predictions support recent experimental findings.
\end{abstract}


\maketitle                   

A major advance in the research on quantum fluids was made with
the recent experimental observation of quantum degeneracy and
off-diagonal long-range order (ODLRO), in a gas of
exciton-polaritons in a semiconductor microcavity
\cite{deng03,kasprzak06,deng06}. Bose-Einstein condensation (BEC)
is the most appealing way of interpreting these findings.
However, the \textcolor {red}{2-D} nature of the system, the
presence of disorder, the hybrid exciton-photon composition of
polaritons, and the composite nature of excitons, call for models
that account for the peculiar aspects of the polariton gas. Along
this line, recent theoretical works have been successful in
describing many specific aspects of the system. Disorder at the
quantum well interfaces, in particular, was accounted for by a
BCS-like theory in which bound excitons were modeled as
particle-hole excitations coupled to the photon field
\cite{keeling04,marchetti06,szymanska06}. For strong exciton
disorder, this theory models the density dependence of the
polariton spectrum and shows how the linear optical response can
probe successfully the phase transition. More generally, in
presence of localization, the lower-energy region of the density
of states is modified with respect to an ideal 2-D system
\cite{kavokin03,keeling06}, making BEC of a trapped polariton gas
the most suited description of the system. Finally, the short
radiative lifetime of polaritons {results in}
non-equilibrium effects
\cite{laussy04,doan05,sarchi06,schwendimann06,szymanska06,wouters07,sarchi07},
leading to condensate depletion \cite{sarchi07} and to a modified
excitation spectrum \cite{szymanska06,wouters07}. These effects
should however be weak at densities above the condensation
threshold \cite{kasprzak06}, and are expected to be negligible
for a polariton lifetime longer than 10 ps.\footnote{Estimated
from a kinetic model~\cite{sarchi07}. For this lifetime, also the
excitation spectrum should tend to the equilibrium one, as can be
seen from Eqs. (8)-(10) of Ref.~\cite{wouters07}.} Very recent
measurements clearly show that thermal equilibrium polariton BEC
can indeed be achieved \cite{deng06}. In spite of the high
relevance of the existing theoretical frameworks, a basic
question remains still unanswered. Are the experimental findings
correctly interpreted in terms of a quantum field theory of
interacting bosons? Such a theory should account self-consistently
for the linear coupling between two Bose fields -- photons and
excitons -- and for the Coulomb and Pauli non-linearities arising
from the composite nature of excitons.

In this Letter, we answer this question by generalizing the
Hartree-Fock-Popov (HFP) \cite{griffin96,shi98} theory of BEC to
the case of two coupled Bose fields. In order to address the
fundamental thermodynamical properties of the polariton gas, we
assume thermal equilibrium -- as was done in previous works based
on non-bosonic models \cite{keeling04,marchetti06} -- and discuss
kinetic effects elsewhere \cite{sarchi07}. {The system, including the Coulomb and Pauli non-linear exciton terms, is described} within an effective
boson Hamiltonian \cite{Rochat00,ben01,Okumura2002},
valid well below the exciton Mott density. We derive coupled
equations for the condensate wave function and the field of
{excitations}, and study the solutions for
parameters {modeling} a recent experiment
\cite{kasprzak06}. We discuss the collective excitation spectrum,
the density-dependent energy shifts, the onset of off-diagonal
long-range order and the phase diagram. Our analysis provides a
generally good account of the experimental findings, in
particular by reproducing the measured critical density and the
energy shifts of the two polariton modes.

Let us introduce the exciton and photon operators $\hat{b}_k$ and
$\hat{c}_k$, obeying Bose commutation rules.\footnote{We assume
scalar exciton and photon fields. The theory can be generalized
to include their vector nature, accounting for light polarization
and exciton spin, as done by Shelykh {\em et al.}, within the
Gross-Pitaevskii limit \cite{shelykh06}.} The polariton
Hamiltonian then reads
\begin{equation}
\hat{H}=\hat{H}_{0}+\hat{H}_{R}+\hat{H}_{x}+\hat{H}_{s},
\label{eq:Hcomp}
\end{equation}
where $\hat{H}_{0}=\sum_{\bf k}\epsilon^{x}_{\bf k}
\hat{b}^{\dagger}_{\bf k}\hat{b}_{\bf k}+\epsilon^{c}_{\bf k}
\hat{c}^{\dagger}_{\bf k}\hat{c}_{\bf k}$ is the non-interacting
term, $\hat{H}_{R}=\hbar\Omega_{R}\sum_{\bf k}(b^{\dagger}_{\bf
k}\hat{c}_{\bf k}+h.c.)$ describes the linear exciton-photon
coupling, and $\hat{H}_{x}=\frac{1}{2}\sum_{kk'q}v_{x}({\bf k},
{\bf k'}, {\bf q})\hat{b}^{\dag}_{{\bf k}+{\bf
q}}\hat{b}^{\dag}_{{\bf k'}-{\bf q}}\hat{b}_{{\bf
k'}}\hat{b}_{\bf k}$ is an effective 2-body exciton Hamiltonian,
modeling both Coulomb interaction and the effect of Pauli
exclusion on electrons and holes. The term
$\hat{H}_{s}=\sum_{kk'q}v_{s}({\bf k}, {\bf k'}, {\bf
q})(\hat{c}^{\dag}_{{\bf k}+{\bf q}}\hat{b}^{\dag}_{{\bf k'}-{\bf
q}}\hat{b}_{{\bf k'}}\hat{b}_{\bf k}+h.c.)$ models the effect of
Pauli exclusion on the exciton oscillator
strength~\cite{Rochat00,Okumura2002}, which decreases for
increasing exciton density \cite{Schmitt-Rink1985}. We account
for the full momentum dependence of $v_{x}({\bf k},{\bf k'},{\bf
q})$ and $v_{s}({\bf k},{\bf k'},{\bf
q})$~\cite{Rochat00,Okumura2002}. They vanish at large momenta,
preventing the ultraviolet divergence typical of a contact
potential~\cite{pita03}.\footnote{In 2-D, many-body correlations
affect significantly the two-body scattering amplitude,
eventually leading to a vanishing $T$-matrix at small collision
energy and in the thermodynamic limit~\cite{lee02}. In the HFP
approximation, it is then customary to replace the interaction
potential $v({\bf k},{\bf k'},{\bf q})$ by the many-body
$T$-matrix $T({\bf k},{\bf k'},{\bf q},E)$ obtained from a
self-consistent summation of ladder diagrams. Here, we have
generalized this approach and computed the many-body $T$-matrices
$T_{x}({\bf k},{\bf k'},{\bf q},E)$ and $T_{s}({\bf k},{\bf
k'},{\bf q},E)$. We find that, for typical parameters, the
correction to $v_{x}$ and $v_{s}$ is of only a few percent.} For
clarity, however, we use the short form $v_{x}({\bf k},{\bf
k'},{\bf q})\rightarrow v_x$ and $v_{s}({\bf k},{\bf k'} ,{\bf
q})\rightarrow\hbar\Omega_R/n_{s}$ in what follows, where $n_s$ is
the saturation density of the exciton oscillator
strength~\cite{Rochat00}.

As non-interacting exciton and photon modes, we assume states in
a square box of area $A$ with periodic boundary conditions. In
real microcavities, interface fluctuations affect the photon
modes, resulting in a disorder potential and in polariton
localization over a few $\mu$m \cite{kasprzak06}. For equilibrium
BEC, the energy spacings between the lowest-lying states
determine locally the effect of quantum fluctuations. Our finite
size assumption thus models in a simple way the local structure
of the spectrum of the disordered system.

Via the Bogolubov ansatz, the exciton and the photon fields are
written as
\begin{equation}
\hat{\Psi}_{x,c}({\bf r},t)=\Phi_{x,c}({\bf
r},t)+\tilde{\psi}_{x,c}({\bf r},t), \label{eq:bogans}
\end{equation}
i.e. as the sum of a classical symmetry-breaking term
$\Phi_{x,c}({\bf r},t)$ for the condensate wave function, and of
a quantum fluctuation field $\tilde{\psi}_{x,c}({\bf r},t)$.
Heisenberg equations of motion result in two coupled equations
for $\Phi_{x,c}({\bf r},t)$. In the HFP approach, anomalous
correlations of the excitation field are neglected
\cite{griffin96,shi98} and we obtain
\begin{eqnarray}
i\hbar\dot{\Phi}_{x}&=&\left[\epsilon^{x}_{0}-2\frac{\hbar\Omega_{R}}{n_{s}}\mbox{Re}\left\{n_{xc}+\tilde{n}_{xc}\right\}+v\left(n_{xx}+\tilde{n}_{xx}\right)\right]\Phi_{x}\nonumber \\
&+&\hbar\Omega_{R}\left(1-\frac{n_{xx}}{n_{s}}\right)\Phi_{c} \nonumber \\
i\hbar\dot{\Phi}_{c}&=&\epsilon^{c}_0 \Phi_{c}+\hbar\Omega_{R}\left(1-\frac{n_{xx}+\tilde{n}_{xx}}{n_{s}}\right)\Phi_{x}.
\label{eq:GPeq}
\end{eqnarray}
Here, we introduce the density matrix
$n_{\chi\xi}=n^{0}_{\chi\xi}+\tilde{n}_{\chi\xi}$
($\chi,\xi=x,c$), where $n_{\chi\xi}^0=\Phi^{*}_{\chi}\Phi_{\xi}$
and $\tilde{n}_{\chi\xi}=\sum_{\bf k\neq 0}n_{\chi\xi}({\bf k})=
\sum_{\bf k\neq 0}\langle \hat{O}_{\chi}^{2}({\bf k})
\hat{O}_{\xi}^{1}({\bf k})\rangle$ are the contributions from the
condensate and from the excited states, respectively. In our
notation, $\hat{O}_{\xi}^{1}({\bf k}) =\hat{O}_{\xi}({\bf k})$
and $\hat{O}_{\xi}^{2}({\bf k})=\hat{O}_{\xi}^{\dagger}(-{\bf
k})$, with $\hat{O}_{x}=\hat{b}$, and $\hat{O}_{c}=\hat{c}$. The
quantities $n_{\chi\xi}({\bf k})$ are computed self-consistently
as described below. By setting $\Phi_{x,c}(t)=e^{-i E t/\hbar}\Phi_{x,c}(0)$ into
(\ref{eq:GPeq}), we obtain a generalized set of two coupled
Gross-Pitaevskii equations for the condensate eigenstate. The two solutions of (\ref{eq:GPeq})
correspond to the lower and upper polariton respectively, and can
be expressed as
${\Phi}_{up(lp)}=X^{up(lp)}_0{\Phi}_{x}+C^{up(lp)}_0{\Phi}_{c}$,
thus fixing the Hopfield coefficients of the polariton
condensate. The low-energy solution $E_0^{lp}=\mu$ defines the
chemical potential.

In the HFP theory, Eqs. (\ref{eq:GPeq}) are coupled to the
field-equations for {excitations}. In analogy with
the case of a single Bose gas, we define the $4\times 4$ matrix
$G({\bf k},i\omega_n)\equiv\{g_{jl}^{\chi\xi}({\bf
k},i\omega_n)\}_{j,l=1,2}^{\chi,\xi=x,c}$, whose elements are the
thermal propagators of the excited particles~\cite{shi98}:
\begin{equation}
g_{jl}^{\chi\xi}({\bf k},i\omega_n)=-\int_{0}^{\beta}d\tau
e^{i\omega_{n}\tau}\langle \hat{O}_{\chi}^{j}\left({\bf
k},\tau\right)\hat{O}_{\xi}^{l}\left({\bf
k},0\right)^{\dagger}\rangle_{\tau}, \label{eq:gij}
\end{equation}
where $\hbar\omega_n=2\pi n/\beta,n=0,\pm 1,...$ are the
Matsubara energies for bosons and $\langle...\rangle_{\tau}$
represents the thermal average of the imaginary-time ordered
product.

The propagator matrix $G({\bf k},i\omega_n)$ obeys the
Dyson-Belaev equation
\begin{equation}
G\left({\bf k},i\omega_n\right)=G^{0}\left({\bf
k},i\omega_n\right)\left[{\bf 1}+\Sigma\left({\bf
k},i\omega_n\right) G\left({\bf k},i\omega_n\right)\right],
\label{eq:dyson}
\end{equation}
where
$G^{0}\equiv \{g^{0}_{jl}({\bf
k},i\omega_n)\}_{jl}^{\chi\xi}=\delta_{\chi\xi}\delta_{jl}[(-)^{j}i\omega_n-\epsilon_{\bf
k}^{(\xi)}+\mu]^{-1}$ is the matrix of the unperturbed propagators, and
\begin{equation}
\Sigma({\bf k},i\omega_n)=\left(\begin{array}{c c}
\Sigma^{xx}({\bf k},i\omega_n) & \Sigma^{xc}({\bf
k},i\omega_n) \\
\Sigma^{cx}({\bf k},i\omega_n) & \Sigma^{cc}({\bf k},i\omega_n)
\end{array}\right) \label{eq:selfen}
\end{equation}
is the $4\times4$ self-energy matrix, here written in a $2\times
2$-block form. In the HFP limit, the {self-energy}
is independent of frequency and reads
\begin{eqnarray}
&&\Sigma_{11}^{xx}=\Sigma_{22}^{xx}=2\left[v n_{xx}-\frac{\hbar
\Omega_{R}}{n_{s}}\left(n_{cx}+n_{xc}\right)\right], \nonumber \\
&&\Sigma_{12}^{xx}=\left(\Sigma_{21}^{xx}\right)^{*}=v
\Phi_{x}^{2}-2\frac{\hbar\Omega_{R}}{n_{s}}\Phi_{x}\Phi_{c}, \nonumber \\
&&\Sigma_{11}^{xc}=\Sigma_{22}^{xc}=\hbar\Omega_{R}\left(1-2\frac{n_{xx}}{n_{s}}\right), \nonumber \\
&&\Sigma_{12}^{xc}=\left(\Sigma_{21}^{xc}\right)^{*}=-\frac{\hbar\Omega_{R}}{n_{s}}\Phi_{x}^{2},
\label{eq:sigmapopov}
\end{eqnarray}
while $\Sigma_{jl}^{cx}=\Sigma_{jl}^{xc}$ and
$\Sigma_{jl}^{cc}=0$. For each value of ${\bf k}$, the analytic
continuation of the Green's functions $g_{jl}^{\chi\xi}({\bf
k},z)$ have four poles at $z=\pm E^{lp(up)}_{{\bf k}}$. They are
the positive and negative eigen-energies of the lower- and
upper-polariton Bogolubov modes. The components of the
corresponding eigenvectors ${\bf h}^{i}_{\sigma}(k)\equiv (X_u,
C_u, X_v, C_v)_{\sigma}^i(k)$ ($i=lp,up$ and $\sigma=+,-$) are the
generalized Hopfield coefficients of the normal $(X_u, C_u)$ and
anomalous $(X_v, C_v)$ kind, with normalization
$|X_u|^2-|X_v|^2+|C_u|^2-|C_v|^2=1$.

The excited-state density matrix $\tilde{n}_{\chi,\xi}({\bf k})$
is related to the normal Green's functions via the relations
\begin{equation}
\tilde{n}_{\chi\xi}({\bf k})=\lim_{\eta\rightarrow 0}\sum_{\omega_n}
e^{i\omega_n\eta}g^{\chi\xi}_{11}({\bf k},i\omega_n).
\label{eq:densgr}
\end{equation}
The corresponding lower- and upper-polariton densities
$\tilde{n}_{lp}({\bf k})$ and $\tilde{n}_{up}({\bf k})$ are
obtained from (\ref{eq:densgr}) in terms of the generalized Hopfield
coefficients. Thus, for a fixed total polariton density $n_{p}$,
the condensed lower-polariton population is given by
\begin{equation}
n_{lp}^0\equiv |\Phi_{lp}|^2=n_{p}-\tilde{n}_{lp}-\tilde{n}_{up}.
\label{eq:condfr}
\end{equation}
From this quantity, $\Phi_x$ and $\Phi_c$ are finally obtained via the Hopfield coefficients for the condensate $X_0$ and $C_0$,
as derived from Eq.(\ref{eq:GPeq}). Hence, a fully self-consistent solution can be obtained by solving iteratively
Eqs. (\ref{eq:GPeq}), (\ref{eq:dyson}), (\ref{eq:densgr}), and (\ref{eq:condfr}), until convergence of
the chemical potential $\mu$ and of the density matrix ${n}_{\chi\xi}({\bf k})$ is
reached. From this solution, for a given polariton density $n_p$ and temperature $T$,
we obtain the spectrum of collective excitations $E_{\bf k}^{lp(up)}$ and the
polariton population $n_{lp(up)}({\bf k})$.

\begin{figure}[ht]
\includegraphics[width=.45 \textwidth]{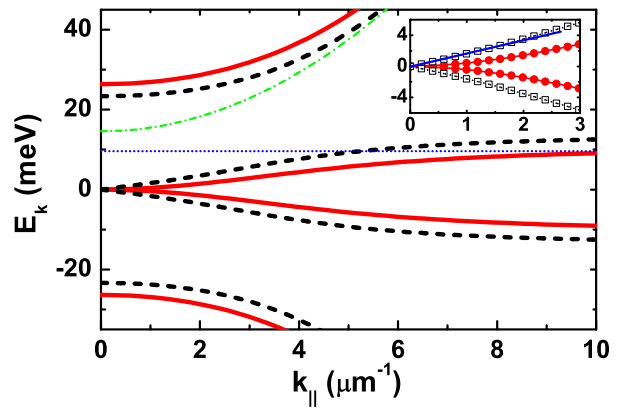}
\caption{The dispersion of the normal modes of the system for
polariton density $n_p=15 \mu m^{-2}$ (solid) and $n_p=400 \mu
m^{-2}$ (dashed). The non-interacting photon (dash-dotted) and
exciton (dotted) modes are also shown. Inset: detail of the
low-energy region, showing the onset of the linear Bogolubov
dispersion (the blue straight line is a guide to the eye).}
\label{fig1}
\end{figure}

For the numerical calculations, we use parameters describing the
experiment in Ref.~\cite{kasprzak06}, i.e. $\hbar\Omega_{R}=13$
meV and detuning $\delta=\epsilon^{c}_0-\epsilon^{x}_0=5$ meV. We
study the results as a function of the system area $A$, ranging
from 100 $\mu \mbox{m}^2$ to 1 $\mbox{cm}^2$. The
momentum-dependent interaction potentials $v_x$ and $v_s$
\cite{Rochat00}, are evaluated for a CdTe quantum well. In Fig.
\ref{fig1} we show the energy-momentum dispersion of the
collective excitations, $\pm E^{lp}_{{\bf k}}$ and $\pm
E^{up}_{{\bf k}}$, as obtained for two different values of the
total polariton density $n_p$ above the condensation threshold,
at $T=20 K$. Close to zero momentum (see inset), the dispersion
of the lower polariton branch becomes linear, giving rise to
phonon-like Bogolubov modes \cite{marchetti06,shelykh06}, as in
the standard single-field theory \cite{pita03}.
{Due to the non-linear terms $\hat{H}_x$ and $\hat{H}_s$, the polariton
splitting decreases for increasing $n_p$.}

\begin{figure}[ht]
\includegraphics[width=.45 \textwidth]{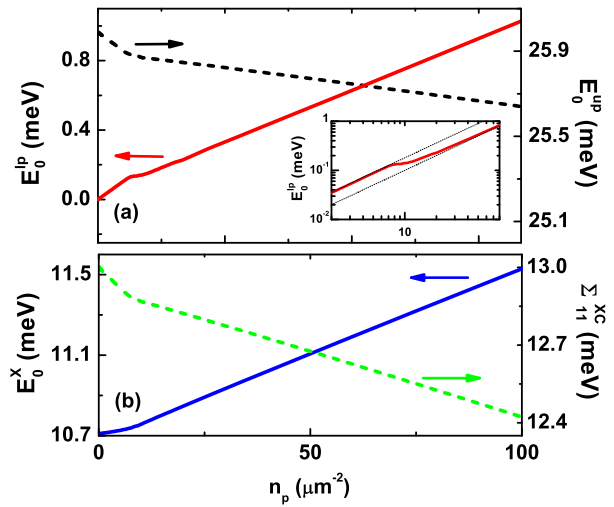}
\caption{(a) Lower (solid) and upper (dashed) polariton energies
at $k=0$ vs polariton density $n_p$. Inset: Double logarythmic
plot of the lower polariton energy (dotted lines: linear slopes
below and above threshold). (b) Bare exciton energy $E_0^{x}$
(solid) and effective exciton-photon coupling $\Sigma_{11}^{xc}$
(dashed). All quantities were computed for $T=20$ K.} \label{fig2}
\end{figure}

The interplay between exciton saturation and interactions in
determining the energy shifts has no counterpart in the BEC of a
single {Bose gas}. Here, the two effects produce
{\em independent} energy shifts of the two polaritons. We plot in
Fig. \ref{fig2} (a) the energy shifts of the two polariton modes
at $k=0$, as a function of the density. Exciton saturation and
interactions result in a {blue-shift} of the lower
polariton and a red-shift of the upper polariton. The shifts are
linear in $n_p$ with the slope changing by a factor of two across
the threshold (see inset), because the contribution of the
thermal populations $\tilde{n}_{xx}$, $\tilde{n}_{xc}$ in Eqs.
(\ref{eq:GPeq}) is twice that of the condensed ones $n^{0}_{xx}$,
$n^0_{xc}$. This trend and the magnitude of the shifts reproduce
fairly well the experimental data \cite{kasprzak06}. To
explain the origin of the opposite shifts of the two polaritons,
we plot in Fig. \ref{fig2} (b) {the exciton energy
$E_0^{x}\equiv \epsilon_0^x+ \Sigma_{11}^{xx}$, and the
exciton-photon coupling $\Sigma_{11}^{xc}$, as a function of
$n_p$. The two quantities contribute comparably to the deviations from the ideal Bose-gas
picture. We predict a very small reduction of the polariton
splitting up to the largest polariton density estimated from the
experiments, thus supporting the idea that polaritons -- as hybrid
exciton-photon quasiparticles -- are stable well above the BEC threshold.}

\begin{figure}[ht]
\includegraphics[width=.42 \textwidth]{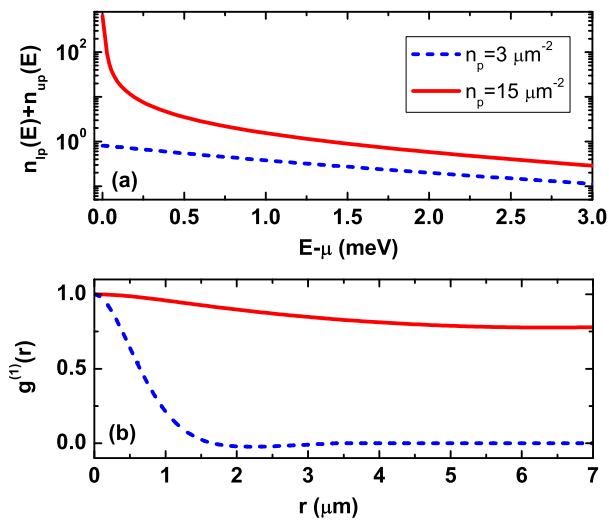}
\caption{(a) Polariton population vs. energy for two values of
$n_p$, computed at $T=20 K$. (b) Corresponding one-body spatial
correlation function of the photon field.} \label{fig3}
\end{figure}

We now turn to study the thermodynamical properties of polariton
BEC. Fig. \ref{fig3}(a) shows the polariton population
$n_{lp}(E)+n_{up}(E)$ for two values of the total density $n_p$,
below and above threshold respectively. {Below
threshold, polaritons follow a Maxwell-Boltzmann distribution.}
Above threshold, the distribution becomes highly degenerate, with
a macroscopic occupation of the lowest-energy state, and a
saturation of the population at high-energy. Fig. \ref{fig3} (b)
displays the simulated one-body spatial correlation function
$g^{(1)}({\bf r})=\langle\hat{\psi}_c^\dagger({\bf
r})\hat{\psi}_c(0)\rangle/(n_{cc}({\bf r})n_{cc}(0))^{1/2}$ of
the photon field. This quantity is directly related, via the
photon fraction, to the actual polariton correlation function,
and models the outcome of an optical experiment.
{It shows the occurrence of ODLRO above threshold
which is the main feature of BEC of an interacting Bose gas
\cite{pita03}.} Below threshold, the correlation extends only
over the thermal wavelength $\lambda_T\simeq1~\mu\mbox{m}$. In
experiments \cite{kasprzak06}, the correlation pattern is shaped
by the interface disorder. However, the measured long-range
correlation is always below 40\%, as compared to 80\% of our
prediction. By means of a kinetic model, we have recently
suggested \cite{sarchi07} that this discrepancy is the main
result of deviations from thermodynamical equilibrium, with
enhanced quantum fluctuations depleting the condensate in favor of
excitations.

\begin{figure}[ht]
\includegraphics[width=.45 \textwidth]{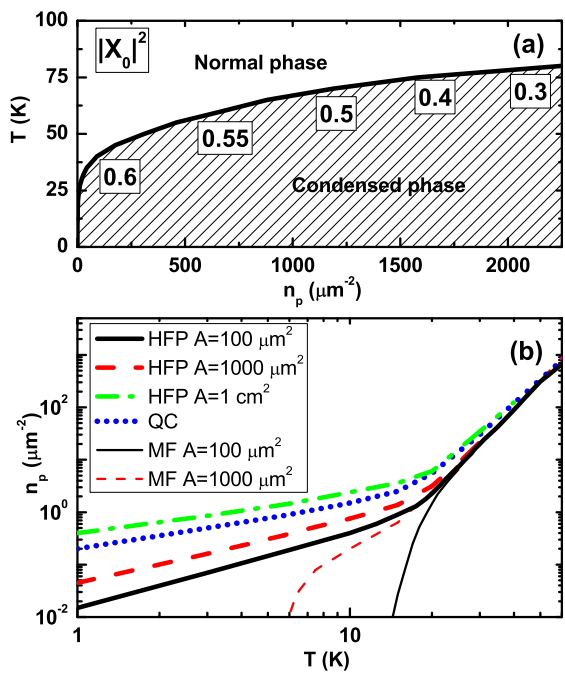}
\caption{(a). Phase diagram of polariton BEC, computed for the
parameters of Ref. \cite{kasprzak06}, and $A=100 \mu \mbox{m}^2$.
The exciton fraction in the condensate $ |X_0|^2$, along the
phase boundary, is indicated in boxes. (b) Detail of the low-$T$
region. HFP denotes the result of the present theory. MF is the
mean-field result. QC denotes the quasi-condensate transition,
corresponding to the onset of a superfluid density.} \label{fig4}
\end{figure}

In Fig. \ref{fig4}(a), we report the computed density-temperature
phase diagram. In the plot, some values of the exciton fraction
$|X_0|^2$ in the polariton condensate are indicated along the
phase boundary. It decreases for increasing density, due to
interactions, but stays finite, {confirming} the
stability of polaritons up to high density. Fig. \ref{fig4}(b)
shows a detail of the low-$T$ region of the phase diagram,
computed for different system areas $A$. {We also
display the physe boundary of the normal-superfluid (quasicondensate)
transition, as obtained from an extension of the Landau formula
\cite{pita03}. For $T<20~\mbox{K}$ and
$A<1000~\mu\mbox{m}^2$, the BEC phase boundary lies well below the quasicondensate
one. Our BEC picture is thus well suited for the description of recent samples \cite{kasprzak06,deng06}, characterized by polariton localization. In the case of a more extended, spatially homogeneous system, a
description in terms of the Berezinski-Kosterlitz-Thouless
transition would be necessary. Fig. \ref{fig4}(b) also shows the
logarithmic increase of the critical density as a function of $A$,
due to the increase of thermal fluctuations. Quantitatively,
the variation is very small, in particular for
$T\geq20~\mbox{K}$ \cite{kasprzak06}. This dependence and the quasicondensate behaviour are hence only expected in samples with improved interface quality and
at lower temperature.} The thin lines in
Fig. \ref{fig4} (b) are the result of a mean-field approximation,
obtained by neglecting the excited states population, i.e.
$n^0_{lp}+n^0_{up}=n_p$. This approximation overestimates the
group velocity at $k=0$, thus strongly underestimating the
critical $n_p$ at low $T$. This points out to the importance of
the HFP approach that we have adopted.

In conclusion, we have generalized the HFP theory to the case of two coupled
Bose fields at thermal equilibrium. The theory allows modeling the BEC of polaritons in
semiconductor microcavities in very close analogy with the BEC of
a weakly interacting gas. The predicted critical density is in good agreement
with a recent measurement~\cite{kasprzak06}. Our analysis thus supports the
interpretation of the experimental findings in terms of a transition to a quantum
degenerate Bose fluid. Open questions remain, basically related to the role of disorder
and localization. If the sample quality can be improved, polaritons will become an
invaluable tool for studying the effects of dimensionality and fluctuations in interacting
Bose systems.

\end{document}